\documentclass[fleqn,usenatbib]{mnras}
\usepackage{amssymb}
\usepackage{varwidth}
\usepackage{graphicx}                   
\usepackage{float}
\usepackage{amsmath}
\usepackage{amssymb}
\usepackage{multirow}
\usepackage{bm}
\usepackage{color}
\usepackage{ulem}
\usepackage[figuresright]{rotating}
\usepackage{appendix}
\usepackage{adjustbox}
\usepackage{txfonts}
\usepackage{ulem}
\usepackage[T1]{fontenc}

\DeclareRobustCommand{\VAN}[3]{#2}
\let\VANthebibliography\thebibliography
\def\thebibliography{\DeclareRobustCommand{\VAN}[3]{##3}\VANthebibliography}

\usepackage{graphicx}   
\usepackage{amsmath}    
\usepackage{amssymb}    

\title[]{An Investigation of the state changes of PSR J2021+4026 and the Vela pulsar}

\author[H.-H. Wang et al.]{
H.-H. Wang$^{1}$\thanks{E-mail: wanghh33@mail.sysu.edu.cn}, J. Takata$^{2}$\thanks{E-mail: takata@hust.edu.cn},  L.C.-C. Lin$^{3}$\thanks{E-mail: lupin@phys.ncku.edu.tw}, P.-H. T. Tam$^{1}$
\\
$^{1}$School of Physics and Astronomy, Sun Yat-sen University, Zhuhai 519000, China\\
$^{2}$Department of Astronomy, School of Physics, Huazhong University of Science and Technology, Wuhan 430074, China\\
$^{3}$Department of Physics, National Cheng Kung University, Tainan 701401, Taiwan}
\date{Accepted XXX. Received YYY; in original form ZZZ}

\begin{document}
\label{firstpage}
\maketitle

\begin{abstract}
{We report on long-term evolution of gamma-ray flux and spin-down rate of two bright gamma-ray pulsars, PSR~J2021+4026 and Vela (PSR~J0835-4510).  PSR~J2021+4026 shows repeated state changes in gamma-ray flux and spin-down rate.  We report two new state changes, a first one from a low gamma-ray flux to a high flux that occurred around MJD~58910, and a second one from high to low flux that occurred around MJD~59510. We find that the flux changes associated with these two state changes are smaller than those determined in the previous events, and 
the waiting time of the new state change from the high gamma-ray flux to low gamma-ray flux is significantly shorter than previous events.
{Since the waiting timescale of the quasi-periodic state changes of PSR~J2021+4026 is similar to the  waiting timescale of the glitch events of the Vela pulsar, we search for the state change of the gamma-ray emission of the Vela pulsar to investigate the possibility that the glitching process is the trigger of the state change of PSR~J2021+4026.} For the Vela pulsar, the flux of the radio pulses briefly decreased around the 2016 glitch, suggesting that the glitch may have affected the structure of the magnetosphere. Nevertheless, we could not find any significant change of the gamma-ray emission properties using 15 years of $Fermi$-LAT data.  {Overall, it seems inconclusive that a glitch-like process similar to that occurred to the Vela pulsar triggers the structure change of the global magnetosphere and causes state changes of PSR~J2021+4026. Further and deep investigations to clarify the mechanism of the mode change for PSR~J2021+4026 are required.}}


\end{abstract}
                     
\section{Introduction}   
Pulsars are highly magnetized and rapidly rotating neutron stars with a stable rotational period. However, there are
two types of timing irregularities: glitches and timing noise. A pulsar glitch is defined as a sudden increase in spin frequency. Although the first glitch was discovered over fifty years ago \citep{Radhakrishnan1969}, it remains an area of active research and interest.

In addition to timing irregularity, there are various intriguing phenomena observed in pulsar's emission, such as mode changes, nulling, intermittency and pulse-shape variability. It has been inferred that some of the phenomena intimately connect with and arise from alterations in the structure of the pulsar's magnetosphere. 
For example, {\cite{lyne2010} report that several mode-changing pulsars, which undergo changes in spin-down states, are accompanied by a clear change in radio pulse profiles}. Mode changes in pulsars are evident through alterations in the pulse profile, encompassing changes in the relative intensity, rotational phase, and widths of individual profile components. These transitions are consistently observed in radio pulsars, as documented by studies such as~\cite{Kramer06}, \cite{WangN2007}, and \cite{shaw22}.
{Mode changes likely signify a clear manifestation of current flow redistribution within the magnetosphere, reflecting alterations in the magnetic field, plasma distribution, or other fundamental elements within this region~\citep{Timokhin2010,Huang2016}}. 

Different timescales and manifestations of mode changes in radio emission and/or the spin-down rate have been observed. For example, PSR J1326–6700 transits between two radio emission states within a timescale of the order of $\sim100$ spin periods~\citep{WangN2007,Wen2020}, while PSR B2035+36 exhibits one abrupt change in spin-down rate on MJD 52950 over 9 years observation, and its radio emission after the mode change is transiting between two states \citep{Kou2018}. 
On the other hand, several mechanisms for mode changes have been observationally and theoretically suggested. For example, connections between the glitching process of neutron stars and mode changes have been observed through radio observations in several pulsars, including PSRs~J1119-6127~\citep{Weltevrede11}, and B2021+51~\citep{liujie2022}, in which a new component in radio emission appears or the pulse profile varies following glitches.  
\citet{Keith13} reported a high correlation between variations in the radio pulse profile and the spin-down rate of PSR B0740-28, which is observed subsequent to the glitch occurring on MJD 55022. \citet{Kou2018} reported a possible glitch event coinciding with a mode change of PSR B2035+36.

As most glitching processes of neutron stars do not show the radiative variation, and some mode changes of pulsars occur without any evidence of accompanying glitches, other physical processes have also been suggested to cause mode changes of the pulsars. 
Precession of neutron stars, which has been suggested as a potential cause for the quasi-periodic changes in the timing residual of PSR~B1828-11, exhibits a mode change lasting $\sim 500$ days in one cycle~\citep{Link2001,Kerr2016,Jones12,Jones2017}. 
The impact of an asteroid has been also suggested to explain a significant decrease in spin-down rate and variation of the radio pulse profile that occurred in 2005 September for PSR J0738-4042 \citep{Brook14}, for which the current spin-down rate still remains lower than the value observed prior to 2005~\citep{Zhousq2023,Lower2023}. \cite{shaw22} mentioned that the mode change of PSR~B2035+36 is similar to that of PSR~J0738-4042. \cite{Basu22} suggested that the evolution of the local magnetic field and the pattern of the sparking discharge in the polar cap acceleration region are causes of the transition between two or more radio emission states.

Variations in emission and mode changes of pulsars have also been observed and investigated with high-energy observations. In particular, since the launch of the $Fermi$-Large Area Telescope (hereafter $Fermi$-LAT) in 2008, the known population of $\gamma$-ray pulsars has significantly increased. The all-sky monitoring of the $Fermi$-LAT enables us to investigate the variability and mode change of the pulsars with the $\gamma$-ray data. 
PSR~J2021+4026 is a radio-quiet pulsar with a spin-period of $P_s\sim 265$~ms, and it is known as the first variable $\gamma$-ray pulsar.  \cite{Allafort2013} reported a sudden decrease in the $\gamma$-ray flux by $\sim 14$~\% and an increase in the spin-down rate by $\Delta\dot{f}/\dot{f}\sim 7$~\%.  Subsequent studies \citep{Ng2016,zhaoj} confirmed that the low $\gamma$-ray flux state continues for about three years and {returns} to the previous high $\gamma$-ray flux state. \cite{Takata2020} found that PSR~J2021+4026 switches between two states with a timescale of several years, namely, a state with high $\gamma$-ray flux and low spin-down rate (i.e., HGF/LSD state) and a state with low $\gamma$-ray flux and high spin-down rate (i.e., LGF/HSD state). 
\cite{Wanghh2018} reported that no significant change in the spectrum and the pulse profile in the X-ray band, which are probably originated from the heated polar cap, has been reported (but see \cite{Razzano23}).  Besides PSR~J2021+4026, \cite{ge2020-1124} reported the model change in the spin-down rate $(\sim 0.4$\%) for PSR~J1124-5916, which is a radio-loud $\gamma$-ray pulsar, using 12 years $Fermi$-LAT data. Although PSR~J1124-5916 is a glitch pulsar, the correlated glitch events at the mode changes were not confirmed in the $Fermi$-LAT data. In X-ray bands, \cite{Marshall15} found a sudden and persistent increase  ($\sim 36$\%)  in the spin-down rate of the radio-quiet pulsar PSR~B0540-69, utilizing data from $RXTE$ and $Swift$ telescopes.

With the timing solution obtained from radio observation or $Fermi$-LAT data,  about 50 $Fermi$-LAT pulsars show glitch activity, and its long-term timing solutions obtained from  $Fermi$-LAT data have been utilized to constrain the glitch mechanisms (e.g. \citealt{gugercinoglu2012}). \cite{Lin2021} investigated the variability of the $\gamma$-ray emission from the glitching pulsar, PSR~J1420-6048, and found the spectral variation of the $\gamma$-ray emission between each glitch. Several $\gamma$-ray pulsars with a characteristic spin-down age similar to that of PSR~J2021+4026 ($\tau_s\sim 7\times 10^4$ years) have displayed glitching activities\footnote{https://www.jb.man.ac.uk/pulsar/glitches/gTable.html\label{atnf}}. Hence, It is not unexpected that PSR~J2021+4026 has also experienced the glitch activities, though such an event has not been confirmed from the long-term timing ephemeris derived from the $Fermi$-data. Since the mechanism of the mode change for PSR~J2021+4026 has not been well understood, further investigation on PSR~J2021+4026 and studies on other glitching $\gamma$-ray pulsars will be desired to examine the influence of the glitch activities on the $\gamma$-ray emission properties.

The Vela pulsar (PSR B0833-45/PSR J0835-4510) is associated with the Vela supernova remnant,  and was discovered over 50 years ago in 1968 \citep{Large1968}.   The Vela pulsar is one of the brightest gamma-ray sources in the sky, with a spin period $\sim 89$~ms and spin-down rate $\sim 1.25\times 10^{-13}~{\rm ss^{-1}}$. It is known as the first pulsar that was observed to undergo glitch~\citep{Radhakrishnan1969} and 24 events have been reported to-date$^1$.
Among all glitching pulsars, the Vela pulsar shows {comparatively frequent glitching activity}
with a waiting time of $\sim$3 years \citep{Dobson2007}, and a relatively large glitch size of $\triangle \nu/\nu\sim 10^{-6}$~\citep{Espinoza11}. Interestingly, \cite{Palfreyman2018} reported a sudden change in the radio pulse shape coincident with the glitch activity on 2016 December 12; an evolution of the pulse profile and polarization was observed in several rotation cycles of the Vela pulsar. The results indicate that the glitch event of the Vela pulsar has an impact on the structure of the magnetosphere. Since the Vela pulsar
is one of the brightest gamma-ray pulsars, it is worthwhile to investigate whether the glitch activity causes any variation or state change in the gamma-ray emission.   

In this paper, we study the temporal evolution of the gamma-ray emission properties for two bright gamma-ray pulsars, PSR~J2021+4026 and Vela pulsar. We {created a} timing solution for each state of {PSR~J2021+4026} and {for each inter-glitch} interval of the Vela pulsar. We compared the pulse profiles and spectra in different time intervals. {Section 2 describes} the data reduction process for the two
pulsars observed by $Fermi$-LAT. In section 3, we present the results of our data analysis and report {a} new state change of PSR~J2021+4026.
In Section 4, we provide a discussion on the implications of our results.

\section{Data reduction}
\subsection{Fermi-LAT {data}}
{$Fermi$-LAT is a  gamma-ray imaging instrument that scans the whole sky approximately every three hours, covering the energy band from $\sim$~20\, MeV to 300\, GeV\citep{atwood09}.} We selected Pass 8 data in the energy band of 0.1-300\, GeV. We choose the position to be R.A.=20$^{h}$21$^{m}$30$^{s}$.48, decl.=40$^{o}$26$^{'}$53$^{''}$.5 for PSR~J2021+4026, and R.A.=08$^{h}$34$^{m}$00$^{s}$.00, decl.=45$^{o}$49$^{'}$48$^{''}$.0 for the Vela pulsar. To avoid Earth's limb contamination, we only included events with zenith angles below 90 degrees. We limited our analysis to events from the point source or Galactic diffuse class (event class = 128) and used data from both the front and back sections of the tracker (evttype = 3).
For our spectral analysis, we constructed a background emission model that incorporated both the Galactic diffuse emission (gll\_iem\_v07) and the isotropic diffuse emission (iso\_P8R3\_SOURCE\_V3\_v1) provided by the $Fermi$ Science Support Center. For each bin of spectra and flux evolution, we refit the data by the binned likelihood analysis (gtlike) and estimate the flux.

We have analyzed data from the $Fermi$-LAT instrument taken between 2008 August and 2023 May to study the gamma-ray emission of PSR J2021+4026. Our analysis focused on phase-averaged spectra, in which we divided the entire dataset spanning from 2008 to 2023 into six inter-glitch intervals: MJD~54710-55850 (high gamma-ray flux/Low spin-down rate state: HGF/LSD~1), MJD~55850-56990 (low gamma-ray flux/high spin-down rate state: LGF/HSD~1), {MJD~56990-58130} (HGF/LSD~2), MJD~58130-58970 (LGF/HSD~2), MJD~58970-59510 (HGF/LSD~3) and $>$MJD~59510 (LGF/HSD~3).
To model the spectrum of PSR J2021+4026, we used a power-law with an exponential cutoff, which can be expressed as:
\begin{eqnarray}
\centering
\label{equa1}
\frac { dN }{ dE }=N_{0} { \left( \frac { E }{ { E_{ o } } }  \right)  }^{ { -\gamma  }_{ 1 } }{ e }^{ -a{ E }^{ { \gamma  }_{ 2 } } }.
\end{eqnarray}   
 The spectral parameters of PSR~J2021+4026 {include} the photon index($\gamma_\mathrm{1}$), $a$ which is related to the cutoff energy, and $\gamma_\mathrm{2}$, the exponent index.
 
To study the evolution of the gamma-ray flux above 0.1 GeV, we divided the data into 60-day time bins. The contribution of each background source is calculated with the spectral parameters obtained from the entire data in each bin. Then, we refit the data by the binned likelihood analysis (gtlike) and estimate the flux.

We also performed the likelihood analysis of {the} Vela pulsar in the time range of 2008 August to 2022 December. We use the power-law with an exponential cutoff model as mentioned above, and the data are divided into seven time segments separated by {glitches}: MJD~54686-55408, MJD~55409-56556, MJD~56557-56922, MJD~56923-57734, MJD~57735-58515, MJD~58516-59417 and  MJD~59418-59900.
By creating timing {ephemeris} for each inter-glitch interval, we study the evolution of the gamma-ray flux and examine the pulse profile in different energy bands. This will allow us to determine whether there are any significant changes in the pulsar's high-energy emission {that is} associated with glitch events.

\begin{figure*}
  \begin{center}
    \centering
    \includegraphics[width=1.8\columnwidth]{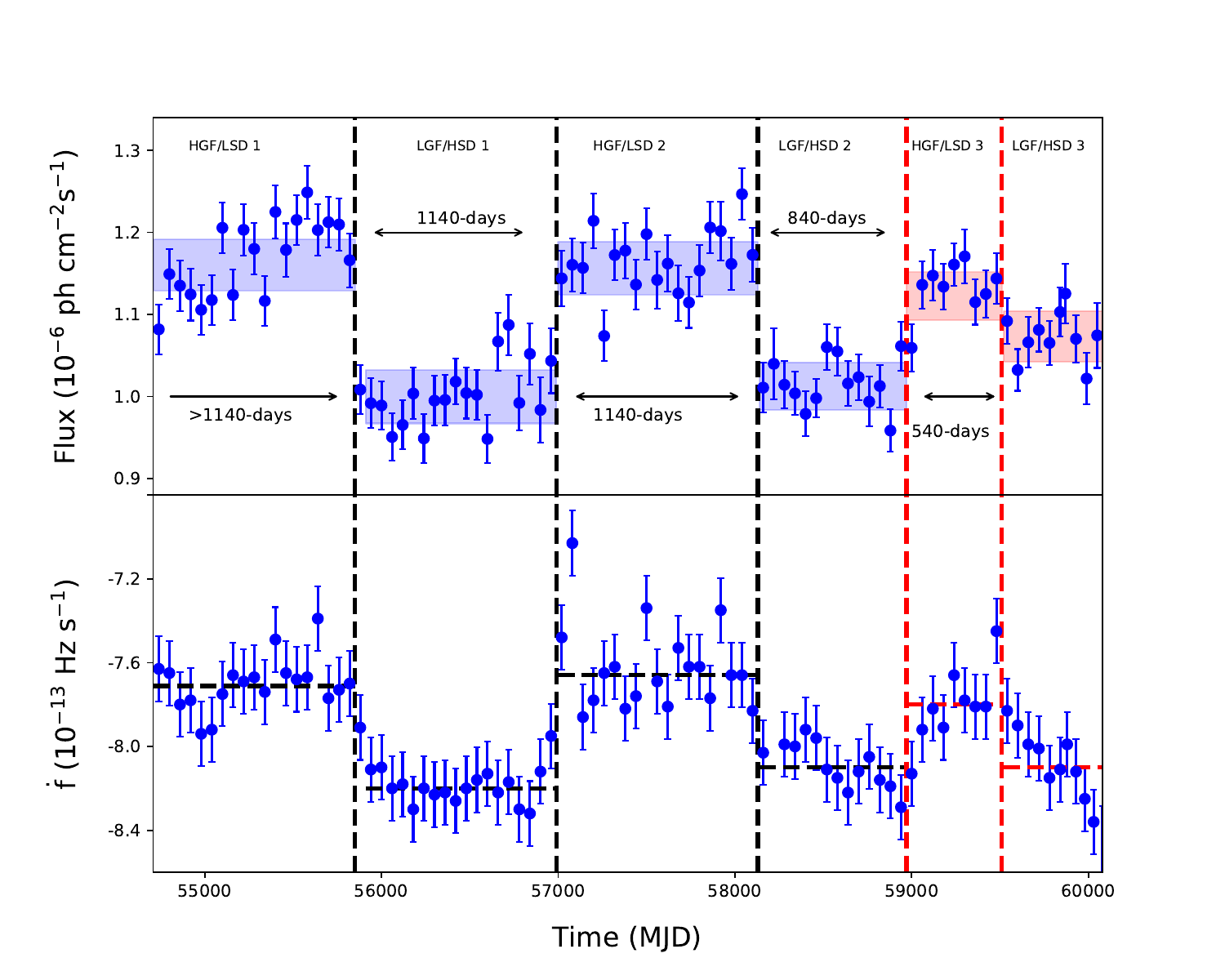}
    \caption{Top: gamma-ray flux ($>0.1$~GeV) evolution of PSR~J2021+4026 from 2008 to 2023. The light-blue/red region labels the uncertainty of the average flux in {different states} in 1~$\sigma$ confidence interval.
    Bottom: The spin-down rate of PSR~J2021+4026 from 2008 to 2023. Each point
 was obtained from the data with a 60-day time bin. The black vertical lines represent the state changes reported in \citet{Allafort2013} and \citet{Takata2020}.
The vertical red dashed lines indicate new state change reported in this study, and the right-most one is determined by the start of the flux drop. The horizontal dashed lines indicate the averaged spin-down rate for each state. }
    \label{gflux2021}
  \end{center}
\end{figure*}

\subsection{Timing analysis}

To ensure that we include the majority of significant source photons, we consider the {photon} events
within a $1^\circ$ aperture centered at the targets. 
We assign phases to these photons using the $Fermi$ plug-in for {TEMPO2 \citep{Hobbs2006,Edwards2006}.} We {barycentred} the photon arrival times to TDB (Barycentric Dynamical Time) using the JPL DE405 Earth ephemeris \citep{standish98}. 
This was done using the "gtbary" task, which provides accurate measurements of the positions and velocities of Solar System.
To obtain the timing ephemeris, we use the Gaussian kernel density estimation method \citep{de1986} provided in \citet{Ray2011} to build an initial template. We then use this template to cross-correlate with the unbinned geocentered data to determine the pulse time-of-arrival (TOA) for each pulse.
Once the pulse TOAs are obtained, we can fit them to an original timing model assumed from the semi-blind search, the timing results of PSR~J2021+4026 and Vela pulsar are shown in Appendix~\ref{appendix}. 

To obtain the temporal evolution of the spin-down rate, we divide the data into several ten-day time
bin and determine the first derivative of the frequency for each time bin.
We also create the long-term ephemeris for the two states of PSR J2021+4026 and for each inter-glitch interval between two glitch events for the Vela pulsar.

\section{RESULTS}

 \subsection{PSR~J2021+4026}

\subsubsection{Temporal evolution}

The top and bottom panels in Figure~\ref{gflux2021} show the flux temporal evolution and first-time derivative of the spin frequency $\dot{f}$, respectively{; each data point is {obtained} with the 60-day $Fermi$-LAT observation. In addition to the previous three events of the state changes reported by \cite{Allafort2013} and
\cite{Takata2020} (vertical black dashed lines in Figure~\ref{gflux2021}), we find new state change 
from LGF/HSD to HGF/LSD around MJD~58910. {\cite{Prokhorov23} also reported the transition to a high-flux state in June 2020 but they did not discuss any changes in the spin-down rate associated with this event} and from HGF/LSD to LGF/HSD around MJD~59510. Since we cannot determine {a precise onset time} of the state change, we indicate {an} approximate location with the red vertical dashed line in Figure~\ref{gflux2021}.} Hereafter, we denote the newly-found states as LGF/HSD~3 and HGF/LSD~3, as indicated in Figure~\ref{gflux2021}.

As Figure~\ref{gflux2021} indicates, the time durations of LGF/HSD~2 and HGF/LSD~3 are shorter than those in the previous states. For example, LGF/HSD~2 continues about 840~days, which is significantly shorter than about 1140~days ($>1140$~days) for HGF/LSD~2 (HGF/LSD~1). We also find that the average flux of the new state is different from the previous values. The flux $\sim 1.12(1)\times 10^{-6}{\rm photon~cm^{-2}~s^{-1}}$ of new HGF/LSD~3 is slightly smaller than $1.15-1.16\times  10^{-6}{\rm photon~cm^{-2}~s^{-1}}$ of the previous HGF/LSD~1 and~2.  The flux level of new LGF/HSD~3 is higher than that in the previous two states of LGF/HSD~1 and~2. The flux changes from HGF/LSD to LGF/HSD in previous two events were $\sim 12-14$\% (ref.~Table~\ref{average-flux}), while the fractional flux change of {the new event} is only $\sim 4$\%. 

A sudden change of the first-time derivative of the frequency, $\dot{f}$, with a time scale of {$<10$}~days was observed at the previous state changes from HGF/LSD to LGF/HSD ~\citep{Allafort2013,Takata2020}. For new event from HGF/LSD~3 to LGF/HSD~3, on the other hand, a sudden change in timing behavior could not be confirmed, and the $\dot{f}$ shows a more gradual evolution to HSD as indicated in the bottom panel of Figure~\ref{gflux2021}. 
{We obtained the spin-down rate at each point with the data of a 60-day time bin.  Neglecting high-order timing noise, we describe the evolution of timing solution using a linear expression as $f(t)=f(t_0 )+\dot{f}(t_0 )(t-t_0 )$, where, $f$ and $\dot{f}$ are spin frequency and frequency derivative, respectively. The middle point of each 60-day time bin is chosen to determine the reference time ($t_0$), the uncertainty is determined from the Fourier width, 1/70 days, and the best frequency and spin-down rate in each bin are determined by the results to gain the most significant detections using H-test~\citep{de2010}}. 
In order to examine the track of the long-term behavior, we calculate the average time derivatives for each state (i.e., horizontal dashed lines in the bottom panel of Figure~\ref{gflux2021}). 
We find that the average value of $\dot{f}=-7.81(5)\times 10^{-13}~{\rm Hz~s^{-1}}$ in new HGF/LSD 3 is smaller than those in previous HGF/LSD states, while $\dot{f}=-8.13(5)\times 10^{-13}~{\rm Hz~s^{-1}}$ in new LGF/HSD~3 
is consistent with the previous LGF/HSD states within errors. 

\begin{figure}
    \centering
    \includegraphics[scale=0.43]{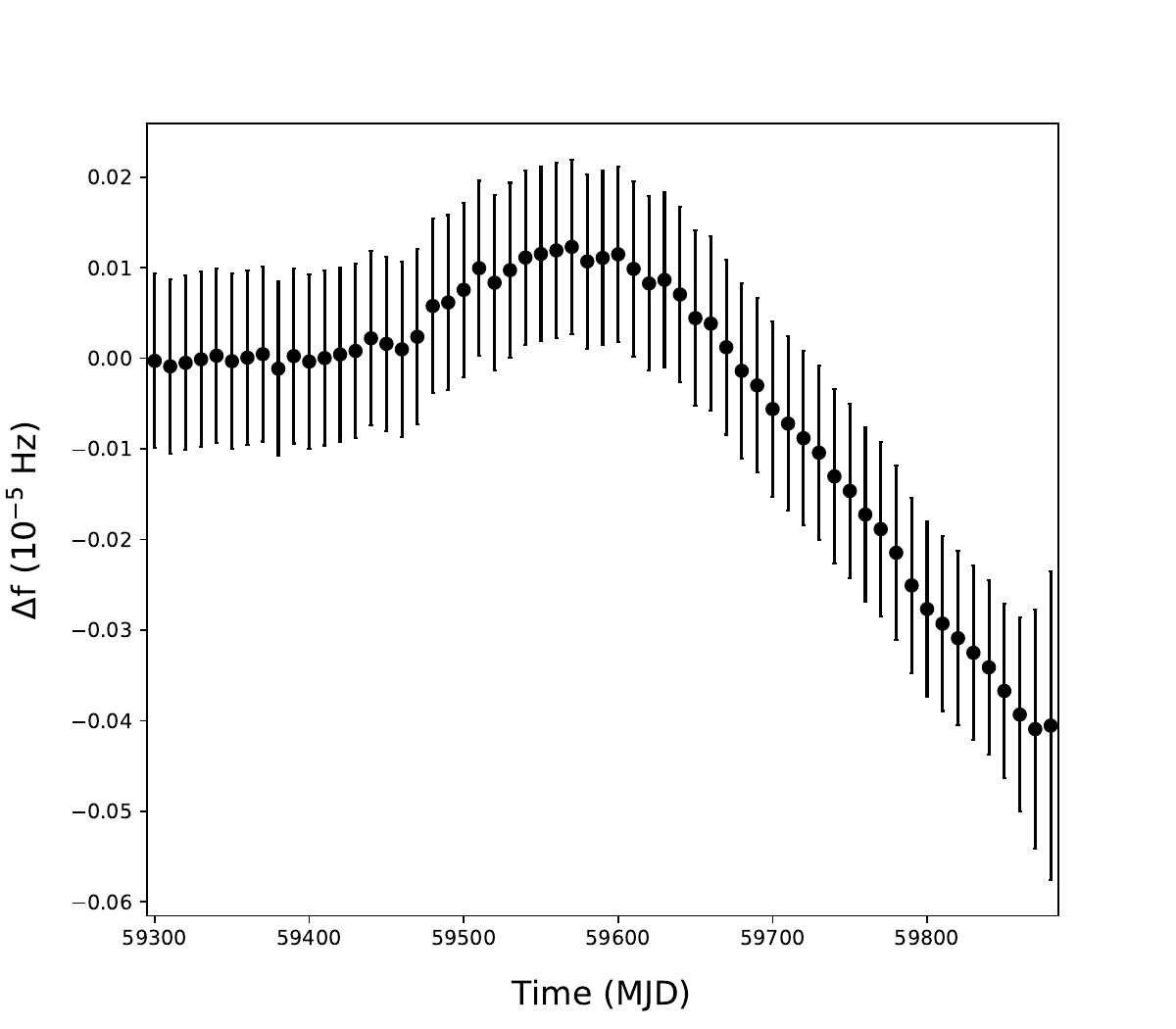}
    \caption{Difference between the observed frequency and predicted frequency from the ephemeris ($\dot{f}$, in the last second column (HGF/LSD~3) in Table~\ref{ephemeris-newMJD}) around MJD~59510. The observed frequency is determined with the data of the 70 day time bin, and two neighbor points overlap the 60-day data. The uncertainty is determined from the Fourier width, 1/70~days.}
    \label{fmodel2021p}
\end{figure}

{Figure~\ref{fmodel2021p}
summarizes the evolution of the spin frequency from HGF/LSD~3 to LGF/HSD 3. {The spin-down rate at each point is calculated with the data of a 70-day time bin, and the starting time difference between two neighbor points is 10 days; namely, two neighbor points overlap the 60-day data.}
In this figure, 
we can see a dramatic change of the spin-down rate at around MJD~59510-59600. 
With the current uncertainty of the data, on the other hand, we cannot confirm the {existence of a glitch.} If there were a glitch, the frequency jump would be smaller than $\triangle f<10^{-7}$~Hz, which is consistent with the previous events from HGF/LSD to LGF/HSD}~\citep{Allafort2013,Takata2020}.

\begin{table*}
   \centering
\caption{Information of the average flux and the parameters of the pulsed structure for PSR~J2021+4026.}
   \label{average-flux}
  \begin{tabular}{ccccccc} 
    \hline
    \hline
   &\multicolumn{2}{c}{2011}&\multicolumn{2}{c}{2018} &\multicolumn{2}{c}{2021} \\
   \hline
     &HGF/LSD~1$^{c}$& LGF/HSD~1$^{d}$ & HGF/LSD~2$^{e}$ & LGF/HSD~2$^{f}$ & HGF/LSD~3 & LGF/HSD~3 
    \\
    \hline 
    MJD       & $<55850$& 55850-56990 & 56990-58130 & 58130-58970 & 58970-59510 & $>59510$ \\
    Flux$^1$  &1.16(1) &0.99(1) &1.15(1) & 1.01(1)& 1.12(1) &1.07(1) \\
     $\triangle F_a^2$ & 0.07(1)& 0.10(1) & 0.06(1) & 0.08(1) & 0.03(1) & 0.02(1) \\ 
    $\triangle F_p^3$ &  & 14(1) & 16(1) & 12(1) & 11(1) & 4(1)\\
    $\dot{f}$ &-7.70(4)& -8.17(4)& -7.63(4) & -8.14(4) & -7.81(5) & -8.13(5) \\
pulse 1$^{a}$ &0.19(2) &0.13(2)& 0.19(2) &0.11(2) &0.16(3) & 0.16(1)\\
pulse 2$^{a}$ &0.176(7)&0.174(6)&0.15(1)&0.16(1)&0.164(1) &0.18(3) \\
pulse~1/pulse~2$^{b}$ &0.54(6)&0.24(3)&0.494(9)&0.26(1) &0.41(1) &0.33(1)\\
\hline
\hline
   \end{tabular}
\begin{flushleft}
~~~~~~~~~~~~~$^1$Average flux in each state ($10^{-6}{\rm photon~cm^{-2}~s^{-1}}$). \\
~~~~~~~~~~~~~$^2$Difference between the average flux of each state and averaged flux of whole data ($10^{-6}{\rm photon~cm^{-2}~s^{-1}}$). \\
~~~~~~~~~~~~~$^3$Flux change from previous state (\%).  \\
~~~~~~~~~~~~~$^{a}$FWHM. \\
~~~~~~~~~~~~~$^b$Ratio of amplitude. \\
~~~~~~~~~~~~~$^c$\cite{Allafort2013}, three-Gaussian components. \\
~~~~~~~~~~~~~$^d$\cite{Allafort2013}, two-Gaussian components.\\
~~~~~~~~~~~~~$^e$\cite{zhaoj}, two-Gaussian components.\\
~~~~~~~~~~~~~$^f$\cite{Takata2020}, two-Gaussian components.\\
\end{flushleft}   
\end{table*}

\begin{table}
   \centering
   \caption{Parameters of Phase-averaged Spectra for PSR~J2021+4026 in two new states.}
   {
  \begin{tabular}{lll} 
    \hline
     & HGF/LSD 3 & LGF/HSD 3 \\
    \hline
    Flux$^1$ &1.74(1)&
    1.69(1)
    \\
    $\gamma_1$ &1.60(3) & 1.74(4)\\
  
   E$_{cutoff}$(GeV) & 2.2(5) &2.4(2)\\
\hline
   \end{tabular}
}
\begin{flushleft}
~~~$^1$ Energy flux of each state in units of $10^{-10}{\rm erg~cm^{-2}~s^{-1}}$. \\
\end{flushleft}   
   \label{2021-all-spec-p}
\end{table}

\subsubsection{Timescale of state change}
In this section, we investigate the correlation between the spin-down rate and the gamma-ray flux of PSR~J2021+4026 from 2008 to 2023 by utilizing the discrete correlation function (DCF). The DCF measures correlation functions without interpolating in the temporal domain \citep{DCF}. The unbinned discrete correlation function is defined as:
\begin{eqnarray}
  \mathrm{UDCF}_{ij} = \frac { (a_{i} -{\overline{a}})(b_{j}-{\overline{b}})}{\sigma_{a}\sigma_{b}},
\end{eqnarray}
where $a_{i}$ and $b_{j}$ {represent} the data lists of gamma-ray flux and spin-down rate, respectively, in
 each time bin of Figure~\ref{gflux2021},  and 
$\overline{a}$ and $\overline{b}$ represent the average values  of the $a_{i}$ and $b_{j}$, respectively. In addition, $\sigma_{a}$ and $\sigma_{b}$ are the standard errors of $a$ and $b$, respectively. 
 We calculate the time-lag of each pair $(a_i, b_j)$ and create the  group of the  pairs for
 every $10^7$~{seconds} of the time-lag ($0<\triangle t_1< 10^7$~s,
 $10^7~{\rm s} <\triangle t_2<2\times 10^7$~s, ...).  For each bin of the time-lag, we obtain number of
 the pairs, $M$,
 and  calculate  the average of UDCF,
\begin{eqnarray}
  \mathrm{DCF}(\triangle t) = \frac{1}{M}\sum \mathrm{UCDF}_{ij},
\end{eqnarray}
where  a standard error for the DCF is defined as,
\begin{eqnarray}
  \sigma_\mathrm{DCF}(\triangle t) = \frac {1}{M-1}{\left\{{\sum{(\mathrm{UDCF}_{ij}-\mathrm{DCF}(t))^{2}}}  \right\}^{1/2}}.
\end{eqnarray}

\begin{figure}
  \includegraphics[width=1.0\columnwidth]{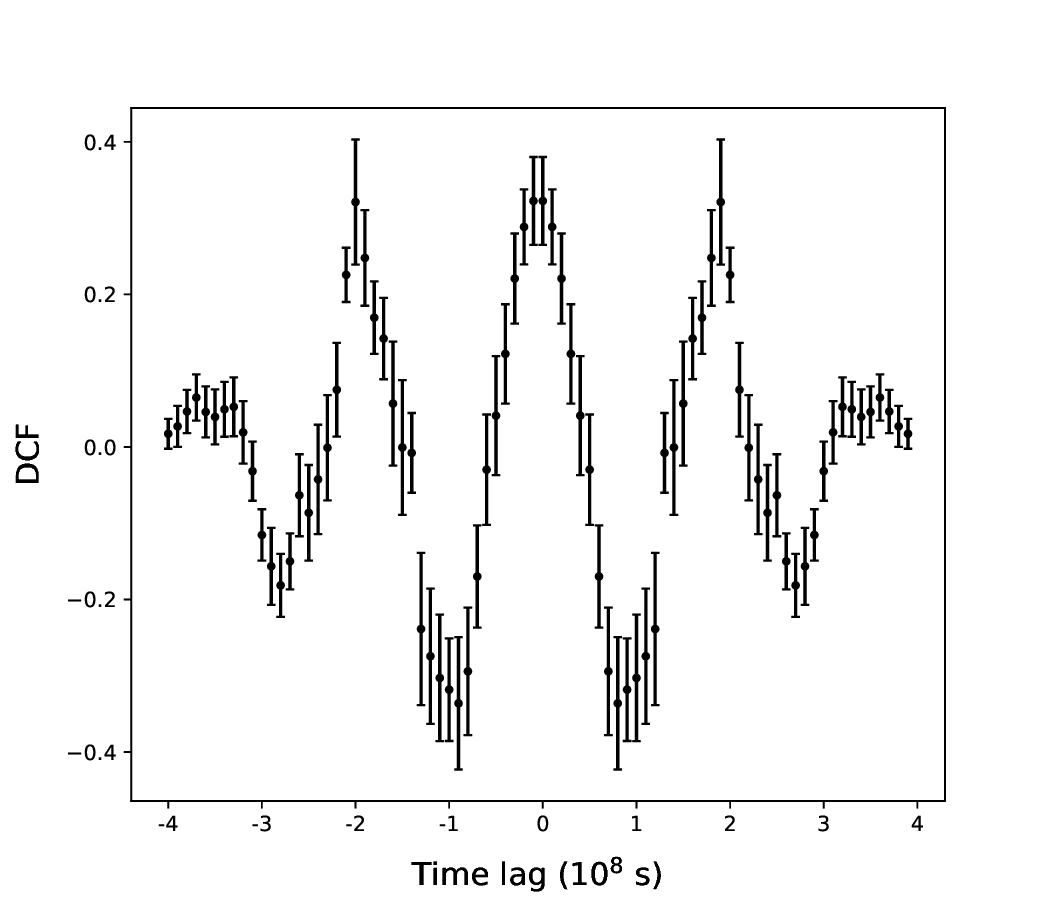}
  \caption{The correlation of $Fermi$-LAT flux and spin down rate of PSR~J2021+4026. The data is from $Fermi$-LAT in the time range of 2008 to 2023.}
  \label{timedcf}
\end{figure}

In Figure~\ref{timedcf}, we present the DCF curve, which exhibits the  maximum correlation coefficient of $\sim$ 0.32 at a time lag of zero {seconds}. This indicates  that the  derivative of frequency changes its state simultaneously with the LAT flux state.
The values of the DCF show  periodicity with a period of 2 $\times$ 10$^{8}$ seconds, which is approximately 6.5 years. This result is consistent with that of \citet{Takata2020}, {which suggests that the time interval between the starting of a specific HGF/LSD is about 6-7 years}.

\subsubsection{Phase-averaged spectrum and pulse profile}
\label{specrum}
\begin{figure}
  \centering
  \includegraphics[scale=0.4]{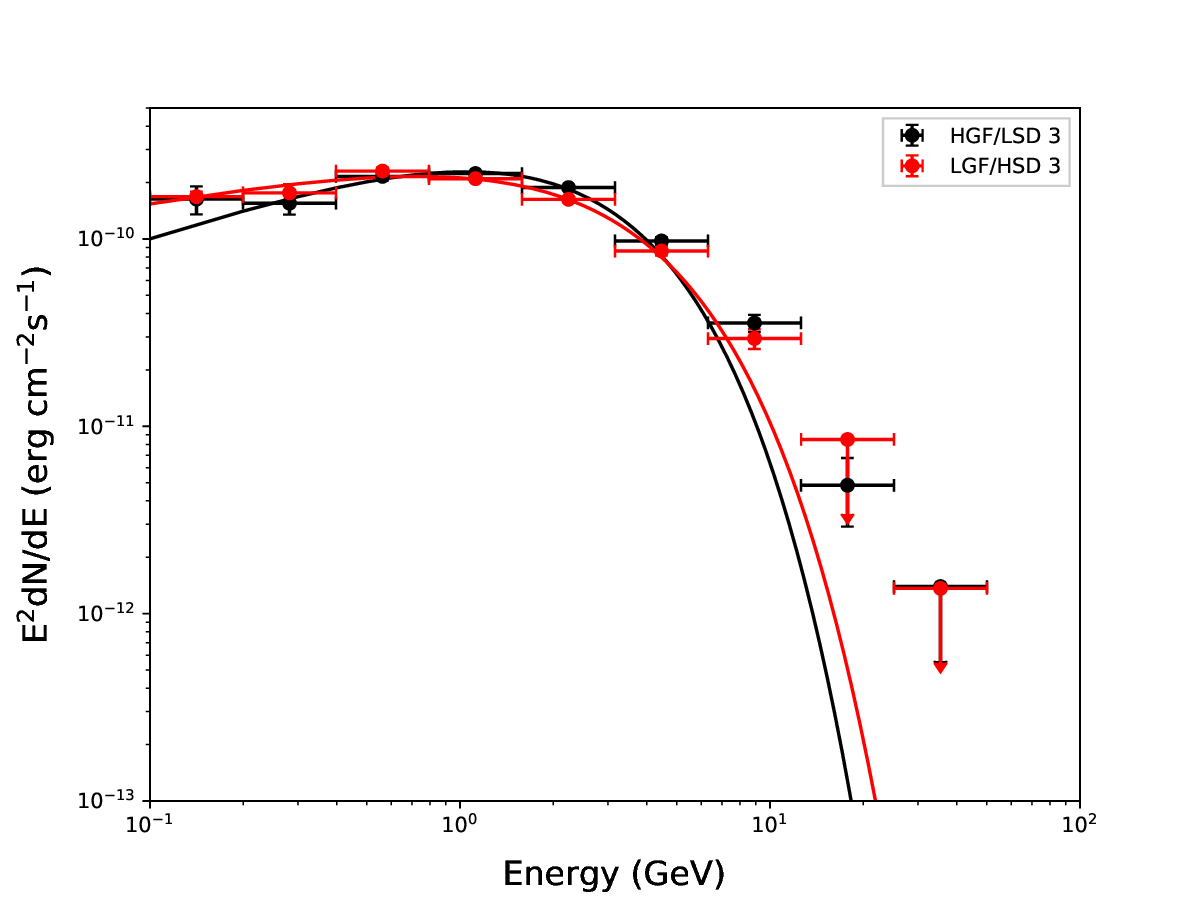}
  \caption{The comparison of spectra at the state of HGF/LSD~3 and LGF/HSD~3. The solid lines show the best fit models for each spectrum according to the ‘PLSuperExpCutoff2’ model (see Equation~\ref{equa1}). }
  \label{compare-2021}
  \end{figure}
In Figure \ref{compare-2021}, we present the observed spectra for the new HGF/LSD~3 (black dots) and LGF/HSD~3 (red dots) states. The solid lines show the fitting function 
using Equation~\ref{equa1}. Table~\ref{2021-all-spec-p} summarizes the parameters of the best fitting function.  As indicated by the timing evolution (Figure~\ref{gflux2021} and Table~\ref{average-flux}), the flux change from HGF/LSD~3 to LGF/HSD~3 is only $\sim 4\%$, which is much smaller than $12-14$\% of the previous two events. The previous state changes from HGF/LSD to LGF/HSD make the spectrum softer; an increase of $\gamma_1$ and a decrease of the cut-off energy were observed. {From the HGF/LSD~3 to LGF/HSD~3, both the power-law index $\gamma_1$ and the cut-off energy increase}.

\begin{figure*}
    \centering
    \includegraphics[scale=0.5]{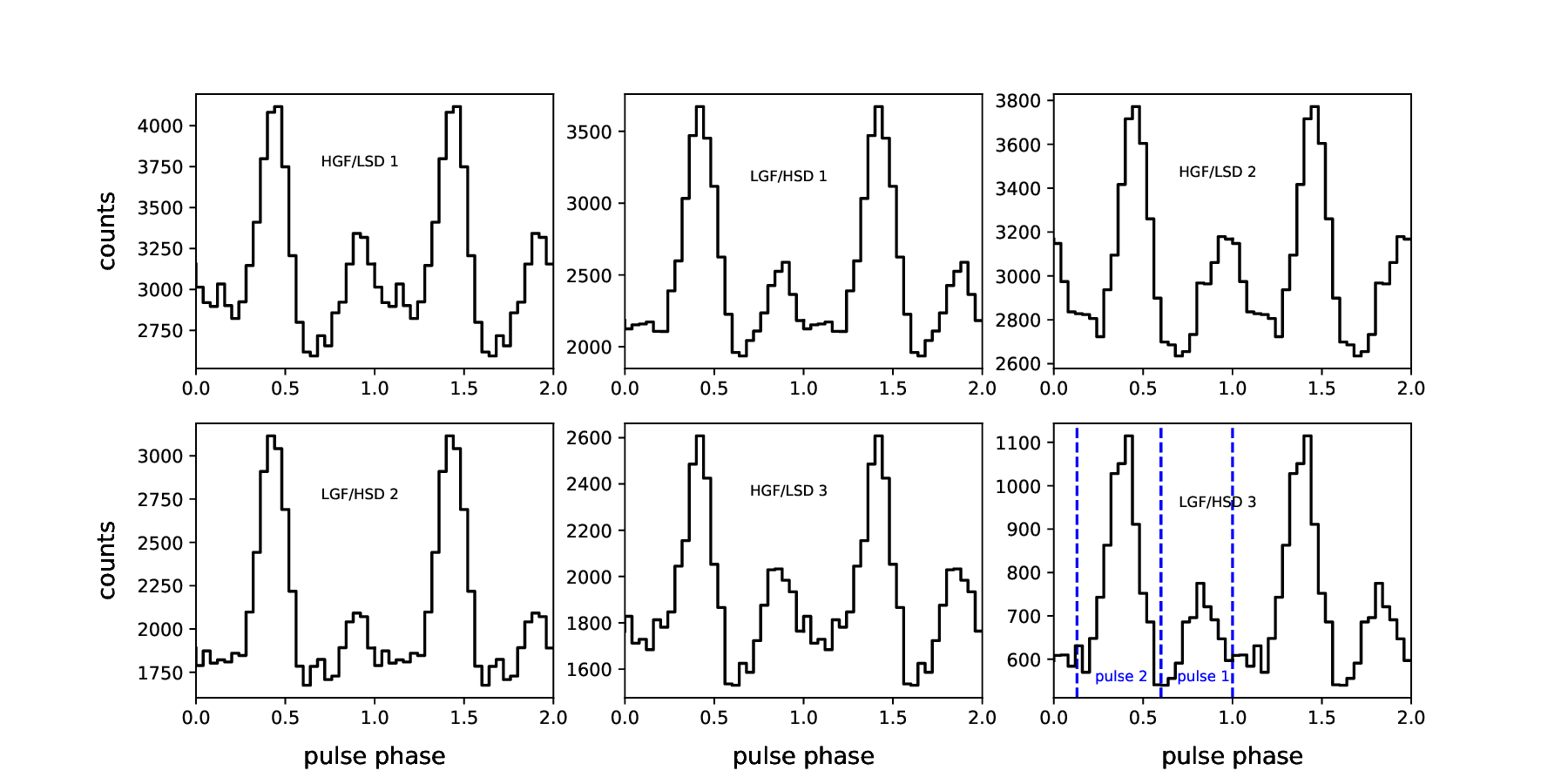}
    \caption{The comparison of the pulse profile for PSR~J2021+4026 at the time of HGF/LSD 1, LGF/HSD 1, HGF/LSD 2, LGF/HSD 2, HGF/LSD 3 and LGF/HSD 3. The pulse profile is generated with photon energy >0.1 GeV. {The vertical dotted lines show the phase interval of pulse~1 and pulse~2}. }
    \label{pulse-2021-all}
\end{figure*}

To investigate the changes in the pulse profile after each state change, we accumulated photons of each new state to derive a global ephemeris for HGF/LSD~3 and LGF/HSD~3 (Table~\ref{ephemeris-newMJD}), and then {folded} the corresponding pulse profiles (as shown in Figure~\ref{pulse-2021-all}). We fit the obtained pulse profiles with two Gaussian functions. Table~\ref{average-flux} summarizes the parameters of the fitting functions for each state. In the previous state changes from HSG/LSD state to LGF/HSD, the ratio of height for the first pulse (small pulse) to the height of the second pulse (main pulse) decreased from about 50\% to 25\%, as Table~\ref{average-flux} indicates. {From HGF/LSD~3 to LGF/HSD~3}, we can see a decrease {in} the ratio of the pulse height, but the magnitude of the change is smaller that in the previous cases. This would be consistent with the smaller flux change in the new event compared to the previous cases.

\subsection{Vela pulsar}
\subsubsection{Long-term behaviors}

\begin{figure*}
   \centering
\includegraphics[scale=0.5]{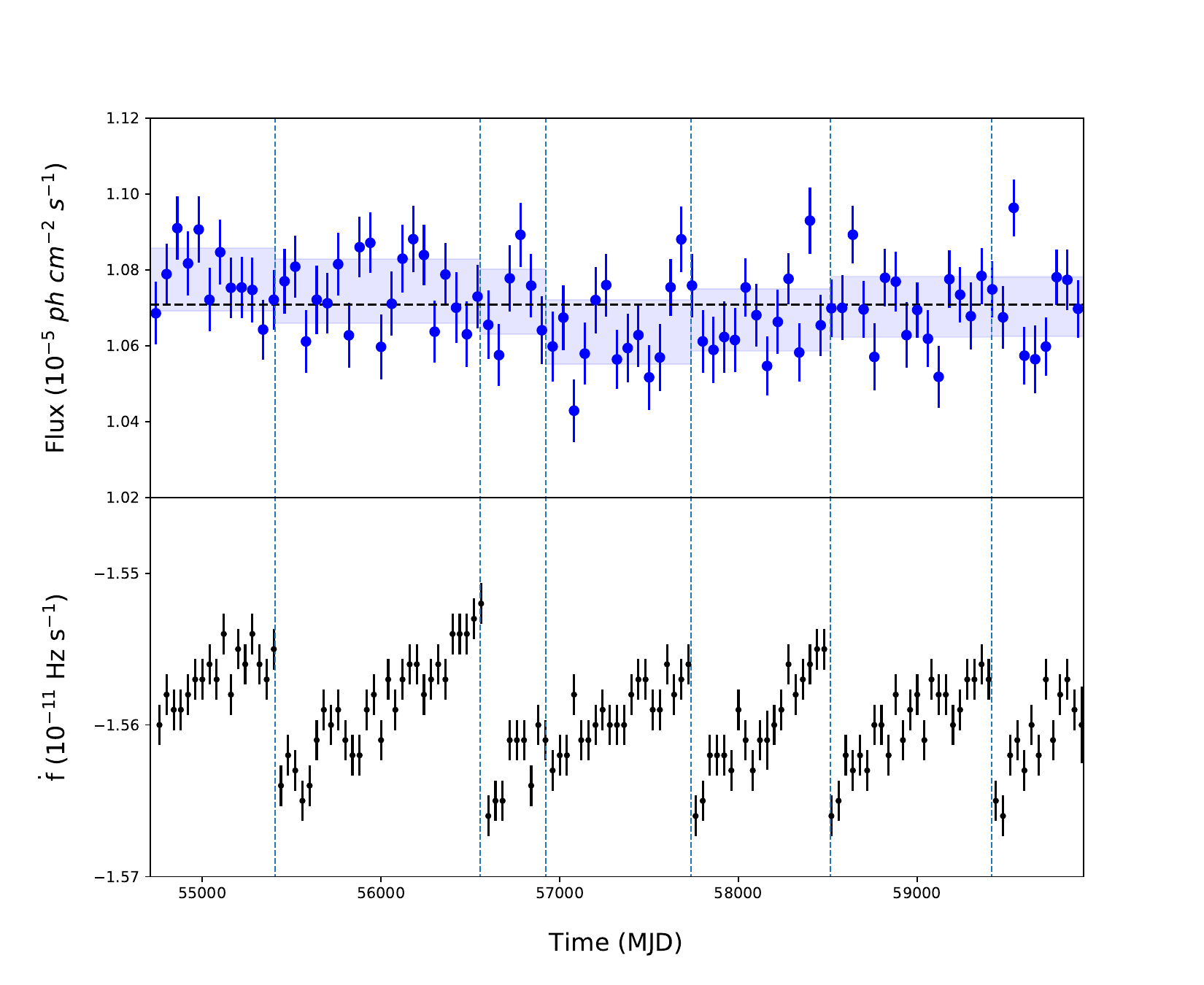} 
   \caption{Evolution of gamma-ray flux (top) and spin-down rate (bottom) of the Vela pulsar. The top panel shows the evolution of gamma-ray flux with a 60-day time bin. The black horizontal line shows the average of flux, the vertical blue lines indicate each glitch epoch, {these glitch epochs are derived from the glitch table\textsuperscript{\ref{atnf}}}. The light-blue region labels the uncertainty of the average flux in each inter-glitch interval in 1~$\sigma$ confidence interval. The bottom panel shows the evolution of spin-down rate, and each data set is obtained from the $Fermi$ telescope with a 40-day time bin.}
   \label{Velaffdot}
  \end{figure*} 

After the launch of the $Fermi$ telescope,  six glitches of the Vela pulsar  have been observed.
We search for the {semi-permanent} state changes of the GeV emission triggered by glitches. Figure~\ref{Velaffdot} presents the evolution of the spin-down rate and the gamma-ray flux of the Vela pulsar. 
We divide the entire $Fermi$-LAT data into seven time intervals, which are bounded by the time of the glitches. {The top panel of Figure~\ref{Velaffdot}} shows the temporal evolution of the gamma-ray flux, and the bottom panel shows the evolution of the spin-down rate. 
Timing ephemeris and the emission properties for each inter-glitch interval are summarized in Table~\ref{Velaephe} and Table~\ref{vela-spec-p}, respectively. 

Despite the frequent glitches of the Vela pulsar, we do not confirm any significant state change in the spectral properties associated  with the glitch. In the bottom panel of Figure~\ref{Velaffdot},  the flux shows a small fluctuation with an amplitude of several percentages. However, such a fluctuation can be explained by  the random fluctuation observed in the data. Table~\ref{vela-spec-p} summarizes the information of the time-averaged spectrum for each inter-glitch interval.
We can find that the spectral properties of the energy flux, the spectral index ($\gamma_1$) and the cut-off energy in the different {time intervals} are {consistent with} each other within {a limited percentage} of errors.

{In Figure ~\ref{Velaffdot}, we determine the average flux with the 1~$\sigma$ uncertainty for each inter-glitch interval (labeled by the black dashed line and light-blue region).
Since the flux uncertainties of each inter-glitch interval are encompassed within the range of the long-term average flux, we conclude that there is no significant variation of the flux in different inter-glitch intervals. 
We also note that one or two data points in each inter-glitch time interval may exhibit notably high or low fluxes, and it is likely caused by random fluctuations. As indicated by Table~\ref{vela-spec-p}, no significant evolution of spectral properties in different inter-glitch intervals is found.} 

\begin{table*}
\caption{Parameters of spectra of {the} Vela pulsar at different inter-glitch {intervals}.}
\centering
 \begin{tabular}{llllllll} 
 \hline
 time range(MJD) & 54686-55205 &55422-56500&56600-56915& 56925-57730 & 57740-58510 & 58525-59400&59430-59900 \\
  \hline
Flux$^{av}$&1.077(8) &1.074(8) &1.071(8)&1.063(8) &1.066(7)&1.070(8)&1.070(7) \\
 Flux$^1$&9.40(2) &9.49(1)&9.37(3)&9.35(2)&9.4(1) &9.39(1) &9.25(2) \\
 $\gamma_1$&1.20(1)&1.205(8)&1.20(1)&1.206(8)&1.214(9)&1.209(7)&1.20(1) \\
 $a$ &0.0479(6)&0.0484(4)&0.0485(7)&0.0484(4)&0.0476(5)&0.0477(4)&0.0475(5) \\
  
pulse 1$^{a}$ &0.022(1) & 0.023(1) &0.023(1)&0.024(1)&0.022(1)&0.024(2) &0.022(4) 
  \\
pulse 2$^{a}$&0.028(1) & 0.028(1) & 0.0278(7) & 0.027(1) &0.029(3)& 0.0264(8) & 0.0245(8)
\\
pulse~2/pulse~1$^{b}$&0.80(8)&0.73(7) & 0.71(5) &0.72(6) &0.7(1) &0.68(6) &0.6(1)
\\
 \hline
 \end{tabular}
 \begin{flushleft}
~~~$^{av}$ Average flux in each state ($10^{-6}$\,${\rm photon~cm^{-2}~s^{-1}}$). \\
~~~$^1$ Energy flux of each state in units of $10^{-8}$\,${\rm erg~cm^{-2}~s^{-1}}$.  \\
~~~$^a$FWHM. \\
~~~$^b$Ratio of amplitude.
\\
\end{flushleft} 
\label{vela-spec-p}
\end{table*}
 
\begin{table}
\vspace{1cm} 
\caption{Energy, arrival time and source probability of  $>$50~GeV photons with P$_{PSR}$ large than max(P$_{PWN}$, P$_{PGAL}$) 
{for the Vela pulsar}.}
\centering
 \label{Vela-energy-above50GeV}
\begin{tabular}{llll} 
\hline
Energy(GeV) &Time(MJD)&Pulse-phase&P$_{PSR}$
\\
\hline
51.08& 55889.11&0.93&0.98 \\
57.41 &56149.28&0.57&0.59 \\
54.586&56317.21&0.93&0.98 \\
51.01&57092.19&0.91&0.81 \\
87.16&57534.33&0.81&0.68 \\
50.82&58316.77&0.90&0.89 \\
67.08&59754.76&0.92&0.97 \\
\hline
\end{tabular}
\end{table}

\subsubsection{Pulse profile at different time intervals}
\begin{figure*}
    \centering
    \includegraphics[scale=0.4]{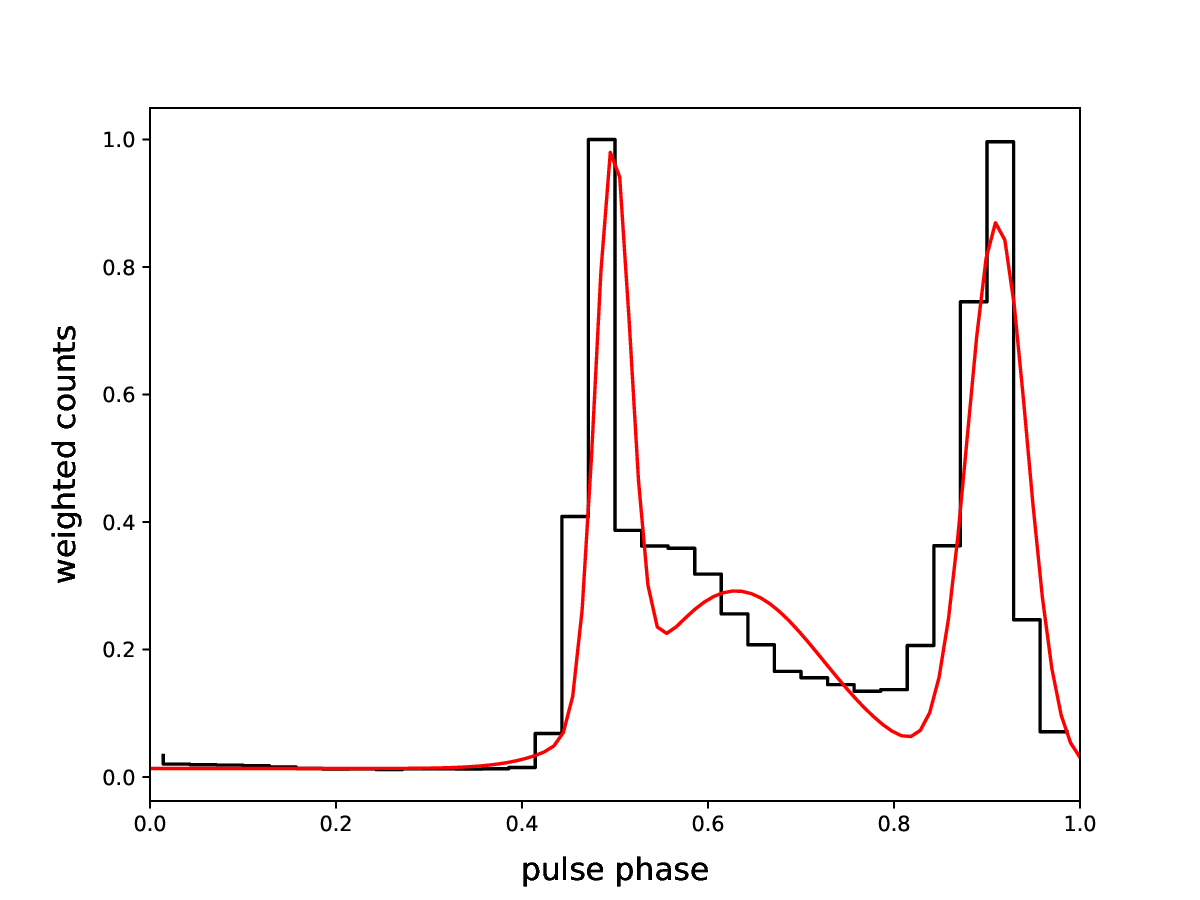}
    \includegraphics[scale=0.4]{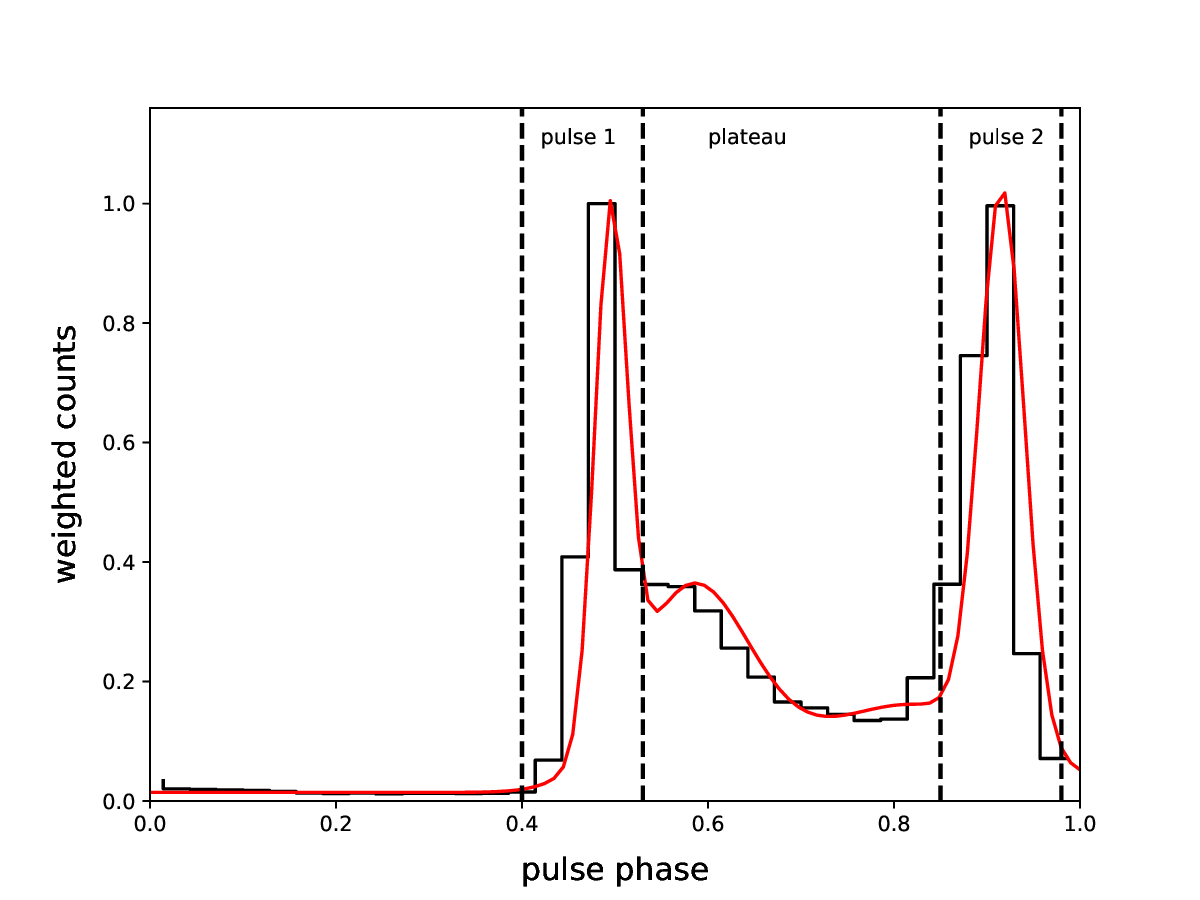}
    \caption{The pulse profile of Vela pulsar (\@MJD 57740-58510) fitted with 3 (left) and 4 (right) Gaussian functions. The black lines show the pulse profile with the weighted photons, while the red lines depict the results of the Gaussian function fitting. The vertical-dashed-balck lines show the phase intervals of pulse 1, pulse 2, and plateau. }
    \label{gauss-3-4}
\end{figure*}

\begin{figure*}
    \centering
    \includegraphics[scale=0.75]{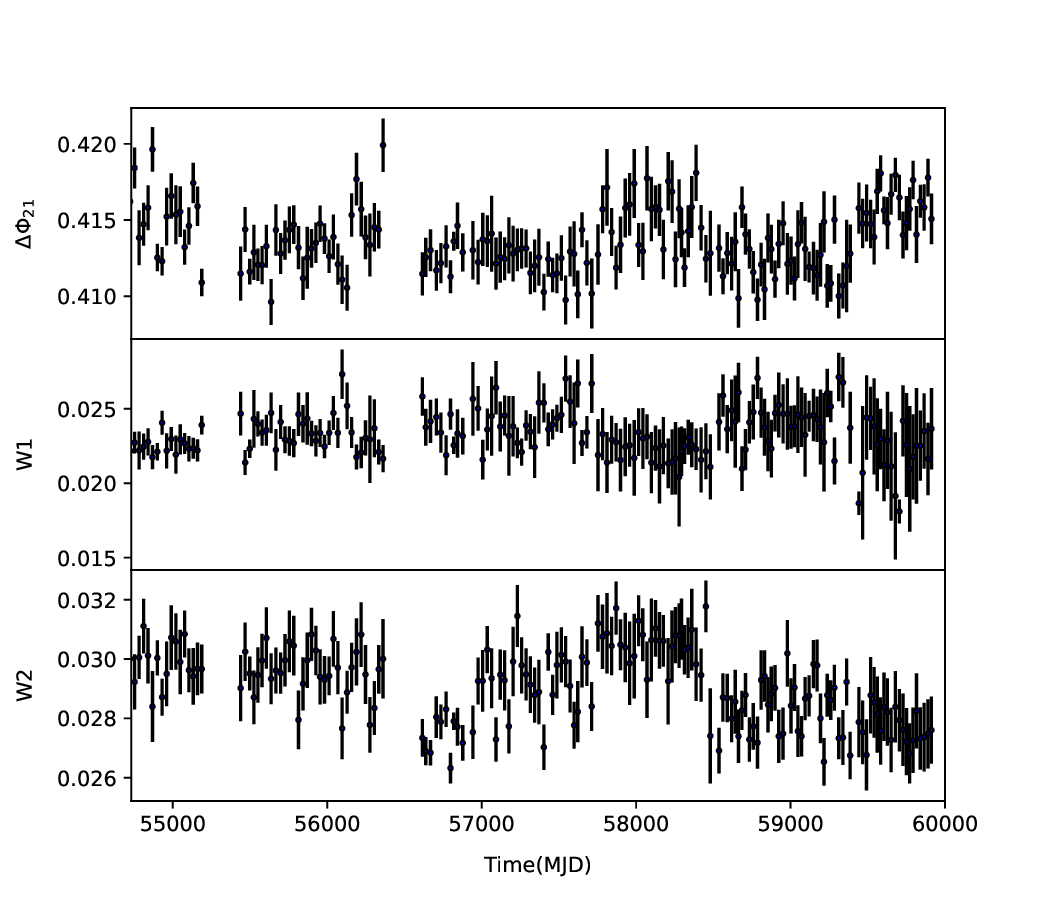}
    \caption{The parameters to describe the pulsed structure in different time intervals of the Vela pulsar. {Each data point} was determined using at least 10000 photons. The upper panel shows the phase separation between {pulse1 and pulse~2} ($\Delta \Phi_{21}$), respectively, as shown in Figure \ref{gauss-3-4}. The middle and the lower panels show the Gaussian width of  {pulse 1 and pulse~2} (W1 and W2), respectively.  {The units of $\Delta \Phi_{21}$, W1 and W2 are the phase in one spin cycle.} 
}
    \label{vela-gauss-epoch}
\end{figure*}

The brightness of gamma-ray emission enables us to investigate the temporal evolution of the {structure} of the pulse profile.
{In order to explore the long-term evolution trend of the pulsed shape, we accumulate $\sim 10^4$ photons in the energy range of 0.1-300~GeV to fold a series of pulse profiles using the ephemeris of each inter-glitch interval, as given in Table~\ref{Velaephe}. 
{We employ a fitting approach on the pulse profile using a combination of four Gaussian functions, which can better fit a} plateau between two main pulses, as illustrated in Figure~\ref{gauss-3-4}. 
{The correlation coefficient of 0.99, derived from the optimal fit employing four Gaussians, surpasses the correlation coefficient of 0.86 obtained when using only three Gaussians to fit the data.}
With a series of the pulse profiles generated with $10^4$ photons, we then determine the temporal evolution of the phase separation between pulse~1 and pulse~2, the Gaussian width of pulse~1 and the Gaussian width of pulse~2 as shown in Figure~\ref{vela-gauss-epoch}.} 
{We find} that there is no significant semi-permanent change shown in the long-term evolution of the pulse profile. 
The glitch of the Vela may still cause some disturbances of the magnetosphere, as indicated in the radio data \citep{Palfreyman2018}. {However, its} effect may be on short-term timescale, or is so small that it is not seen in the gamma-ray emission properties averaged over a time scale much {longer than} 10 days.

\subsubsection{Searching for $>$100~GeV photon with $Fermi$-LAT data}
\cite{leung2014} searched for the pulsed emission above 50~GeV using about 5.2~years {of} $Fermi$-LAT data and reported 5 photons {that probably originated} from the Vela pulsar. {They also report the highest energy photon about 208~GeV with a significant level of 2.2~$\sigma$}. The results of H.E.S.S.-II \citep{HESS2018} also suggest that the spectrum of the pulsed emission from the Vela pulsar extends beyond 100~GeV.

We revisit the search for very {high-energy gamma-ray photons} with $Fermi$-LAT data using updated  $Fermi$-LAT gamma-ray catalog (PASS~8 data),
galactic diffuse emission (gll\_iem\_v07.fits) and isotropic background emission (P8R3\_SOURCE\_V3) models, while \cite{leung2014} used
the reprocessed Pass 7 "Source" class (P7REP\_SOURCE\_V15 IRFs). Following \cite{leung2014}, firstly, we divide the $Fermi$-LAT data obtained
within MJD 54710-60000 into 50-300~GeV with a ROI (region of interest) of 4-degree in radius. To select photons originating from the pulsar, we
used the $gtsrcprob$ tool to calculate the probability of each photon within ROI associated with the pulsar. We also calculated the probabilities of the photons coming
from the Vela X (P$_{PWN}$) and the Galactic diffuse emission (P$_{GAL}$). We selected only photons with P$_{Vela}$ $>$ max(P$_{PWN}$, P$_{GAL}$). Subsequently, the
same methodology was applied to each state, utilizing the respective ephemeris obtained from the timing analysis.

\begin{figure}
  \begin{center}
    \centering
    \includegraphics[scale=0.38]{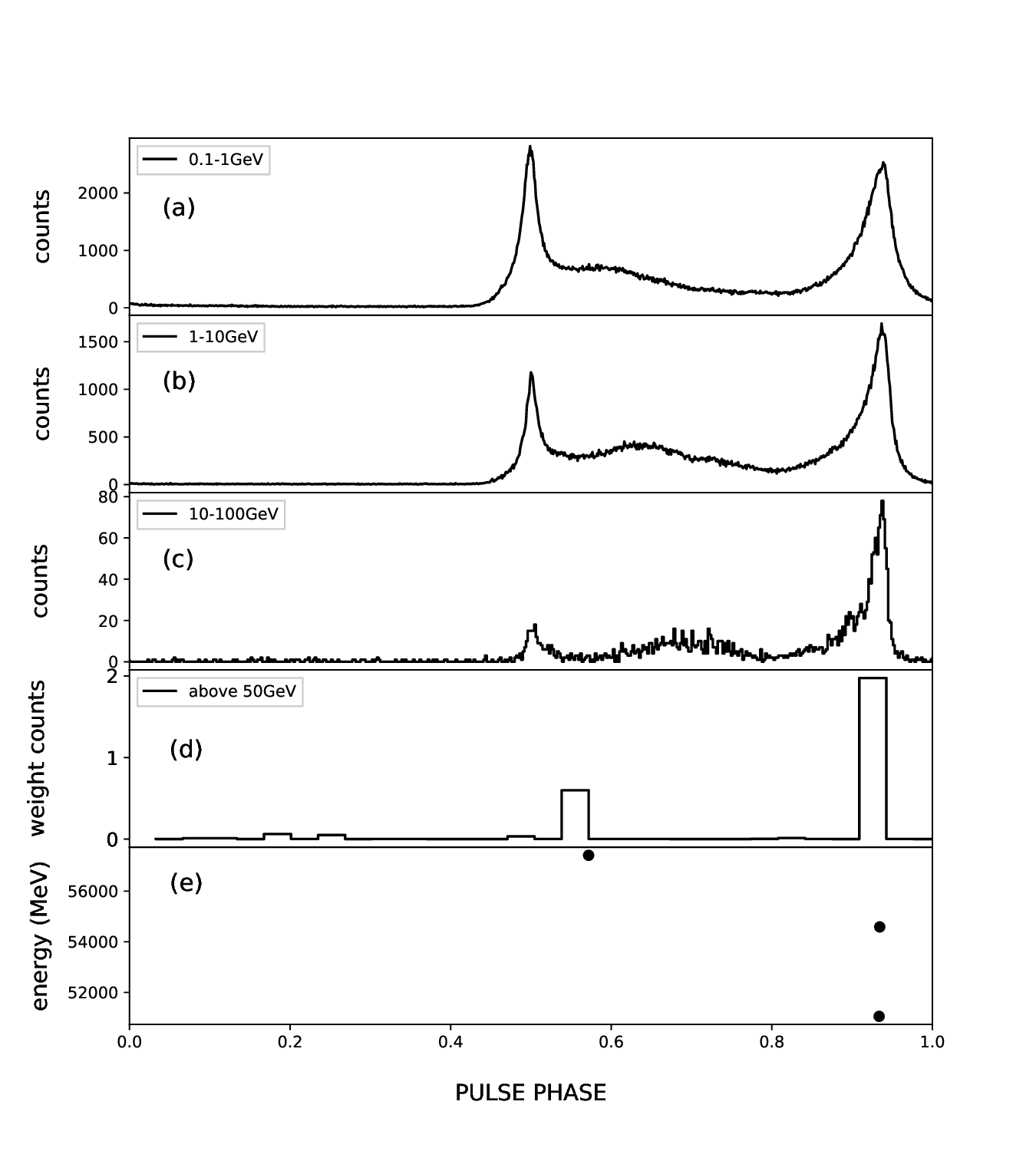}  
\caption{{From (a) to (c): the folded light curve in the energy range of 0.1-1~GeV, 1-10~GeV, and 10-100~GeV using  the photons collected within $1^{\circ}$ radius centered at the source. From (d) to (e): the weighted light curve in the energy of 50-300~GeV and the energy distribution of the photons above 50~GeV, which are collected within 
$4^{\circ}$ radius and with a source probability of  $P_{Vela}$ $>$ max($P_{PWN}, P_{GAL}$) in the same way as \citet{leung2014} . The data collected during MJD~54686-56583 as an example to find the energy of photons.} } 
    \label{Velaall-lc}
  \end{center}
\end{figure}

 Figure \ref{Velaall-lc} shows the light curve in different energy bands from panel (a) to (c), the panel (d) shows the weighted light curve above 50~GeV.  
 The results in the MJD~54686-56583 in comparison to \citet{leung2014} are depicted in panel (e) of Figure \ref{Velaall-lc}, which illustrates the energy distribution of photons. We detect 7 photons with an energy larger than 50~GeV shown
in Table \ref{Vela-energy-above50GeV}. {The arrival times of two photons, observed at MJD 56149 and MJD 56317, are consistent with the results reported by~\cite{leung2014}. However, the} energy of the photon detected at MJD~56317, $\sim 54.6$~GeV, is
significantly smaller than $\sim 79.5$~GeV of the previous result.  Although we employ the same method as \citet{leung2014} to investigate the emission in different time ranges, we do not confirm any photons above 100 GeV.

\section{Discussion and summary}
We have investigated the temporal evolution of the characteristics of the GeV emission from two young pulsars, PSR J2021+4026 and the Vela pulsar. For PSR J2021+4026, {we confirm new state changes from LGF/HSD to HGF/LSD and from HGF/LSD to LGF/HSD}, but the magnitude of the change in the gamma-ray flux and the spin-down rate is smaller than those seen in previous state changes. {The pulse shape change} is also small compared to previous events. Concerning the state change of the emission and spin-down properties, we speculate that the change of global electric current circulating
the magnetosphere could lead to corresponding state change of the emission and spin-down properties, and a larger (smaller) electric current produces larger (smaller) spin-down rate in LGF/HSD (HGF/LSD) state.	
We expect that the change of gamma-ray flux is due to the change of the size of the particle acceleration/emission regions, and HGF/LSD (LGF/HSD) has a larger (smaller) size of the gap. {There is a tendency in the previous states that the HGF/LSD state} has a larger cut-off energy in the spectrum and  the width of the pulse profile
is wider \citep{Takata2020}. {This tendency can be understood if the HGF/LSD has a larger acceleration/emission region}. Although it is expected that the magnitude of the electric current and the size of	the acceleration/emission region {are} related~\citep{Hirotani2006},  a positive or negative correlation  will depend on the location of the acceleration/emission
region and the geometry of the magnetosphere \citep{takata2006}.  {For the most recent state change from HGF/LSD to LGF/HSD}, the small change in the spin-down rate implies the change of the electric current is smaller, which results in a small change in the structure of the acceleration/emission region.

{\cite{Takata2020} proposed that the crust cracking process triggers the mode change of PSR~J2021+4026, and it causes a sudden change in the magnetic field structure of the polar cap region. According to this model, the timescale between each mode change from HGF/LSD state to LGF/HSD state  corresponds to the timescale to accumulate the magnetic stress inside the crust, and the waiting timescale in LGF/HSD state is the relaxation timescale during which the polar cap structure returns to its original state.}  

{\cite{Jones12} suggested a possible correlation between the neutron star's spin period and the waiting timescale of mode change of the transient radio pulsar, and argued the idea of free precession of neutron stars with strained crusts. In this model, the capacity of the acceleration of particles within the magnetosphere was linked to different phases of precession. 
The precession model has been widely applied to the observed periodic variations of the spin-down rate of PSR B1828-11~\citep{Link2001,Ashton2017}.  For PSR J2021+4026, we found that the waiting time in LGF/HSD~2 ($\sim 840$~days) and HGF/LSDF~3 ($\sim 530$~days) were shorter than the previous corresponding states~($\sim 1140$~days), indicating that the mode change is quasi-periodic event.  Such a quasi-periodicity may be incompatible with the free-precession model, for which the stable periodic variation of the spin-down rate will be expected. }

{Among the transitioning radio pulsars, PSRs~J0738-4042 and B2035+36 have shown a mode change with relatively large variations in the spin-down rate, $|\Delta\dot{f}|/f\sim 7$\% and 12\%, respectively. The magnitude of the spin-down rate change of the two pulsars is similar to  $|\Delta\dot{f}|/f\sim 7$\% of PSR~J2021+4026. Moreover, the waiting time to the next mode change of three pulsars will be of the order of or longer than $\sim 10$~years; it is observed once or twice mode changes of PSR~J0738-4042 over 50 years~\citep{Lower2023} and once of PSR~B2035+36 over 9 years~\citep{Kou2018}. \cite{Brook14} argued the mode change of PSR~J0738-4042 with an asteroid encounter model~\citep{Cordes08}. In this model, the asteroid with neutral charge can approach to the pulsar magnetosphere, and it is evaporated and ionized by an X-ray irradiation of the pulsar. The ionized gas that migrates within the magnetosphere changes the torque exerted on the neutron star and the emission properties. Due to the size of the frequency jump and waiting time in several years, the asteroid encounter may be an attractive model as the origin of the mode change of PSR~J2021+4026. 
}

{
Since about 50 glitching $Fermi$-LAT pulsars have been confirmed$^1$, it is not unexpected that PSR~J2021+4026 has experienced the glitch process of the neutron star. In this paper, we focus on the Vela pulsar, which is both a glitching and bright $\gamma$-ray pulsar, we analyzed $Fermi$-LAT data of the Vela pulsar to investigate the influence of glitch on its $\gamma$-ray emission.} \cite{Palfreyman2018} reported sudden changes in the radio pulse shape coincident with the 2016 glitch event of Vela pulsar, indicating the glitch affected the structure of the magnetosphere.  \cite{bransgrove2020} suggested that the glitch launched the Alfven wave, which subsequently {caused} the high-energy radiation and the electron-positron pair creation. They suggest that the created pairs quench the radio emission at the 2016 event of the Vela pulsar. Although the enhancement of the gamma-ray flux is expected during propagation of the Alfven wave in the magnetosphere, the predicted timescale of the existence of the wave, $\sim 0.2$~s, is too short to be investigated with $Fermi$-LAT data. We also did not find any evidence for the change of the magnetosphere structure {for the Vela pulsar} in a timescale of years, as discussed in section~3.2. {Consequently, we cannot conclude that the glitch is the trigger of PSR~J2021+4026 based on the current study.}

In summary, we have examined the evolution of gamma-ray flux and spin-down rate in two bright gamma-ray pulsars, PSR~J2021+4026 and the Vela pulsar. 
{We confirmed new states change from LGF/HSD to HGF/LSD and from HGF/LSD to LGF/HSD observed for PSR~J2021+4026, occurring around MJD~58910 and MJD~59510, respectively.} We found that  the waiting time, the change in the flux and the first time derivative of frequency are smaller than previous events. The change {in} the  shape of the pulse profile is also smaller than those of the previous events.
Our results confirm that the state change of PSR~J2021+4026 is not regularly repeated but quasi-periodically repeated with the different magnitude of change of the gamma-ray flux and the spin-down rate. 
We also carried out the timing and spectral analysis of the Vela pulsar to investigate the effect of the glitch on the observed gamma-ray emission properties, since
the radio observation indicates that the glitch disturbs the structure of the magnetosphere. However, we did not confirm any significant state change of
the gamma-ray emission triggered by the glitch of the Vela pulsar.  
Finally, using the 15-year $Fermi$-LAT data of the Vela pulsar to search photons above~100~GeV, we did not find any photons above 100~GeV from the $Fermi$-LAT data.
We still did not find any evidence for the change of the magnetosphere structure in a timescale of years in the 2016 glitch, while the created pairs may quench the radio emission.

\section*{Acknowledgements}
We express our appreciation to an anonymous referee for useful comments and suggestions.
We appreciate Drs Kisaka and S.Q. Zhou for useful  discussion on glitching pulsars. This work made use of data supplied by the LAT data server of the Fermi Science Support Center (FSSC) and the archival data server of NASA’s High Energy Astrophysics Science Archive Research Center (HEASARC). H.H.W. is supported by the Scientific Research Foundation of Hunan Provincial Education Department (21C0343). J.T. is supported by National Key Research and Development Program of China ( 2020YFC2201400) and the National Natural Science Foundation of China (grant No. 12173014).
L.C.-C.L. is supported by NSTC through grants 110-2112-M-006-006-MY3 and 112-2811-M-006-019. H.H.W. and P.H.T. are supported by the National Natural Science Foundation of China (NSFC) grant 12273122 and a science research grant from the China Manned Space Project (No. CMS-CSST-2021-B11).

\section*{Data Availability}
(i) The Fermi-LAT data used in this article are available in the LAT data server at https://fermi.gsfc.nasa.gov/ssc/data/ access/. \\
(ii) The Fermi-LAT data analysis software is available at https: //fermi.gsfc.nasa.gov/ssc/data/analysis/software/. \\
(iii) We agree to share data derived in this article on reasonable request to the corresponding author.

\bibliographystyle{mnras}
\bibliography{example} 
\newpage
\appendix
\section{Ephemeris of PSR J2021+4026 and Vela pulsar}
\label{appendix}
{We performed the timing analysis in HGF/LSD~3 and LGF/HSD~3 of PSR~J2021+4026 and each state of Vela pulsar}. We use the Gaussian kernel density estimation method to build an initial {template}. We then use this template to cross-correlate with the unbinned
geocentered data to determine the pulse time-of-arrival (TOA) for
each pulse. {We then fitted the TOAs according to a timing model to obtain the spin frequency, spin-down rate and other parameters}. 
In order to further {investigate} the emission behavior between different states, we extract the source events within a 1$^o$ radius centered at the target. As shown in Table~\ref{ephemeris-newMJD} and Table~\ref{Velaephe}, {ephemeris} of each state are  obtained to compare the shapes of pulse profile of each state.

\begin{table*}
 \caption{Ephemerides of PSR~J2021+4026 derived from LAT Data {covering} MJD~54710-60000.}
   \label{ephemeris-newMJD}
   \centering
   \begin{tabular}{lllllll}
   \hline
     Parameters & & \\
     \hline
     Right Ascension &  \multicolumn{6}{c}{20:21:29.99} \\
     Declination &  \multicolumn{6}{c}{+40:26:45.13} \\
     \hline
     Valid MJD range&54710-55850&55850-56980&56990-58130&58140-58900& 58920-59500&59550-60000\\
     Pulse frequency,$f(s^{-1})$ &3.769066845(7)&3.768952744(6)&3.768891254(7)&3.768823695(3)
     &3.768775022(1)&3.768734495(7)\\
     1st derivative,$\dot{f}(10^{-13} s^{-2})$&-7.762(1)&-8.206(1) &-7.699(1)&-8.132(1)
     &-7.822(4)&-8.173(3)\\
     2nd derivative,$\ddot{f}(10^{-21} s^{-3})$&0.51(3)&0.56(2) &-0.12(1)&1.98(3)
     &2.19(1)&-13.2(1)\\
     3rd derivative,$\dddot{f}(10^{-28} s^{-4})$&-0.33(4)&69.3(3)&-0.31(4)&3.6(1)
     &0.4(2)&169(6)\\
     4th derivative,$f^{(4)}(10^{-35} s^{-5})$&-0.11(1)&-1.14(7) &0.38(5)&-8.3(3)
     &-5.6(8)&-94(2)\\
    5th derivative,$f^{(5)}(10^{-41} s^{-6})$&0.13(3)&-0.22(1)&0.058(1)&-1.26(9)
    &0.3(1)1&281(9) \\
    6th derivative,$f^{(6)}(10^{-48} s^{-7})$&-0.26(5)&0.11(1)&-0.08(1)&2.9(3)
    &-1.8(4)&-316(9) \\
    7th derivative, $f^{(7)}$(10$^{-55}$\,s$^{-8}$)&
    0.23(3)&0.51(3)&-0.06(1)&2.8(6)
    &$\cdots$& 51(9) \\
8th derivative, $f^{(8)}$(10$^{-62}$\,s$^{-9}$)&-0.08(1)&0.33(2)&0.10(2)&-9(2)
&$\cdots$&210(3)
\\
    Epoch zero (MJD)&54936&56600&57500&58500&59200&59800   \\
    Time system & \multicolumn{6}{c}{TDB(DE405)} \\
     \hline
   \end{tabular}
\end{table*}
\newpage

\begin{table}
\caption{The ephemerides of Vela before and after each glitch.}
\centering
\rotatebox[origin=c]{90}{
\begin{varwidth}{\textheight}
 \label{Velaephe}
\begin{tabular}{lllllllll} 
\hline
Parameters & & & & & 
\\
\hline
Right Ascension &  \multicolumn{5}{c}{08:35:20.6033089} 
\\
 Declination & \multicolumn{5}{c}{-45:10:34.82304 } 
\\
\hline
Valid MJD range & 54686-55205 &55422-56500&56600-56915& 56925-57730 & 57740-58510 & 58525-59400&59430-59900
\\
Spin frequency, $f$(s$^{-1}$) & 
11.19025957434(1) & 11.189275117(8)&11.18782702(7)
&11.186479678(1) &11.1860906325(4) & 11.18531013(1)&11.18411112(1)
\\
1st derivative, $\dot{f}$(10$^{-11}$\,s$^{-2}$) &-1.55759513(7) &-1.56292(1) &-1.5614(4)&
-1.55703(1) & -1.561418(2)&-1.56366(8)&-1.5633(1)
\\
2nd derivative, $\ddot{f}$(10$^{-21}$\,s$^{-3}$) &0.791(2) &1.88(3)&
21(2) &0.86(1) & 1.690(7) & -1.56366(8)&-6.6(6)
\\
3rd derivative, $\dddot{f}$(10$^{-28}$\,s$^{-4}$ & -0.7509(3)&0.17(1)&
-16(2)&0.091(4)& -0.202(7)&-3.1(4)& 0.60(6)
\\
4th derivative, $f^{(4)}(10^{-34}$\,s$^{-5})$ & 0.042(3)&-0.13(9)&
-6.9(1)& $\cdots$ & $\cdots$ &2.9(4)&5.2(4)
\\
5th derivative, $f^{(5)}$(10$^{-41}$\,s$^{-6}$) & 0.481(1)&-0.15(8)&13(2) &$\cdots$ &  $\cdots$ &-6.6(1)& -0.10(1)
\\
6th derivative, $f^{(6)}$(10$^{-48}$\,s$^{-7}$) & -0.26(4)&0.7(1) & -5.4(1)&$\cdots$ &  $\cdots$ & 7(1) & 6.7(8)
\\ 
7th derivative, $f^{(7)}$(10$^{-55}$\,s$^{-8}$) & -2.71(5)&-0.9(2)&$\cdots$& $\cdots$& $\cdots$ &-5.1(9)&  $\cdots$
\\
8th derivative, $f^{(8)}$(10$^{-63}$\,s$^{-9}$) & -1.2(2) &4(1)& $\cdots$&$\cdots$& $\cdots$ & 14(2) & $\cdots$
\\
Epoch zero (MJD) &54853&55600 &56700&57700 &58000&58600&59500
\\
Time system & \multicolumn{7}{c}{TDB(DE405)} \\
\hline
\end{tabular}
\end{varwidth}}
\begin{flushleft}
{\small
~~~~~~~Notes. The numbers in parentheses denote 1$\sigma$ errors in the last digit.\\
}
\end{flushleft}

\end{table}


%

\bsp	
\label{lastpage}
\end{document}